\documentclass{aa}  

\usepackage[varg]{txfonts}
\usepackage[pdfpagelabels=false]{hyperref}
\hypersetup{colorlinks=true,
            linkcolor=blue,
            citecolor=blue,
            filecolor=blue,
            urlcolor=blue,
            }
    
\usepackage{graphicx}	        
\usepackage{amsmath, bm}        
\usepackage{amssymb}	        
\usepackage{multirow}           
\usepackage{subfig}

\newcommand{\hkpc}{{\ifmmode{h^{-1}{\rm {kpc}}}\else{$h^{-1}{\rm{kpc}}$}\fi}}
\newcommand{\hMsun}{{\ifmmode{h^{-1}{\rm {M_{\odot}}}}\else{$h^{-1}{\rm{M_{\odot}}}$}\fi}}
\newcommand{\Msun}{{\ifmmode{{\rm {M_{\odot}}}}\else{${\rm{M_{\odot}}}$}\fi}}

\newcommand{\Sec}[1]{Section~\ref{#1}}

\newcommand{\Fig}[1]{Fig.~\ref{#1}}

\newcommand{\code}[1]{\texttt{#1}}

\begin{document}

   \title{The Three Hundred Project: The stellar angular momentum evolution of cluster galaxies}

   \author{R. Mostoghiu\inst{1,3,4}
          \and
          A. Knebe\inst{1,2,3}
          \and
          F. R. Pearce\inst{4}
          \and
          C. Power\inst{3,5}
          \and
          C. D. P. Lagos\inst{3,5}
          \and
          W. Cui\inst{6}
          \and
          S. Borgani\inst{7,8,9,10}
          \and
          K. Dolag\inst{11,12}
          \and
          G. Murante\inst{8}
          \and
          G. Yepes\inst{1,2}
          }

   \institute{Departamento de F\'isica Te\'{o}rica, M\'{o}dulo 15, Facultad de Ciencias, 
              Universidad Aut\'{o}noma de Madrid, E-28049 Madrid, Spain\\
              \email{robert.mostoghiu@uam.es}
         \and
             Centro de Investigaci\'{o}n Avanzada en F\'isica Fundamental (CIAFF), Facultad de Ciencias, 
             Universidad Aut\'{o}noma de Madrid, 28049 Madrid, Spain
         \and
             International Centre for Radio Astronomy Research, 
             University of Western Australia, 35 Stirling Highway, Crawley, Western Australia 6009, Australia
         \and
             School of Physics \& Astronomy, 
             University of Nottingham, Nottingham NG7 2RD, UK
         \and 
             ARC Centre of Excellence for All Sky Astrophysics in 3 Dimensions (ASTRO 3D)
         \and
             Institute for Astronomy, 
             University of Edinburgh, Royal Observatory, Edinburgh EH9 3HJ, United Kingdom
         \and
             Dipartimento di Fisica, Sezione di Astronomia, 
             Universit\`{a} di Trieste, via Tiepolo 11, I-34143 Trieste, Italy
         \and
            INAF - Osservatorio Astronomico Trieste, via Tiepolo 11, 34123, Trieste, Italy,
         \and
            Institute of Fundamental Physics of the Universe, via Beirut 2, 34151 Grignano, Trieste, Italy
         \and
            INFN, Instituto Nazionale di Fisica Nucleare, Trieste, Italy
         \and
            University Observatory Munich, Scheinerstra{\ss}e 1, D-81679 Munich, Germany
         \and
            Max-Planck-Institut fur Astrophysik (MPA), 
            Karl-Schwarzschild Stra{\ss}e 1, D-85748 Garching bei M\"{u}nchen, Germany
             }

   \date{Received -- / Accepted -- }


   \abstract{Using 324 numerically modelled galaxy clusters as provided by \textsc{The Three Hundred} project, we study the evolution of the kinematic properties of the stellar component of haloes on first infall. We select objects with M$_{\rm{star}}>5\times10^{10}\hMsun$ within $3R_{200}$ of the main cluster halo at $z=0$ and follow their progenitors. We find that although haloes are stripped of their dark matter and gas after entering the main cluster halo, there is practically no change in their stellar kinematics. For the vast majority of our `galaxies' -- defined as the central stellar component found within the haloes that form our sample --  their kinematic properties, as described by the fraction of ordered rotation, and their position in the specific stellar angular momentum$-$stellar mass plane $j_{\rm star}$ -- M$_{\rm star}$, are mostly unchanged by the influence of the central host cluster. However, for a small number of infalling galaxies, stellar mergers and encounters with remnant stellar cores close to the centre of the main cluster, particularly during pericentre passage, are able to spin-up their stellar component by $z=0$.}

   \keywords{methods: numerical -- 
            clusters: general -- 
            galaxies: evolution -- 
            galaxies: kinematics and dynamics
            }
               
   \titlerunning{The stellar angular momentum of cluster galaxies}
   \authorrunning{Mostoghiu et al.}
   \maketitle
   

 
\section{Introduction} \label{sec:introduction}
    In a hierarchical model of structure formation, the structures observed today are a result of the merging of dark matter clumps at high redshift via gravitational collapse. As the clumped dark matter grows into larger objects to form haloes, they experience tidal torques from neighbouring objects. At the same time, baryons condense in the centre of such structures to form galaxies \citep{Peebles69,Doroshkevich70,White78,White84}. During their mutual formation both the dark matter and baryonic component experience the same tidal fields, and hence it is expected that they gain the same amount of specific angular momentum. Furthermore, considering that baryons evolve inside dark matter haloes, it is also expected that the kinematic evolution of galaxies is influenced by the halo in which they reside. 
    
    However, galaxies and their haloes also feel the influence of the environment. Haloes falling towards and eventually orbiting within galaxy clusters are disrupted by a series of processes predominant in the cluster environment, e.g. ram-pressure stripping \citep{Gunn72,Abadi99,Bahe15,Arthur19,Mostoghiu21} that removes the gas in haloes and quenches the star formation of galaxies; galaxy harassment \citep{Moore96,Moore98,Smith10,Smith15},  mergers \citep{Dressler80,Hashimoto00,Behroozi14}, tidal torques \citep{Fujita98,Balogh00,Park07}, and interactions in general \citep{Knebe06a,Recchi14} that can disrupt the haloes' components; and dynamical friction \citep{Valtonen90,Jiang00,Fujii06,vanDenBosch17,Miller20} which slows down infalling haloes and, over time, causes them to fall to the centre of the cluster. While the stars residing deep inside the potential well of the halo are shielded from tidal effects, they might nevertheless feel and react to the change of its own halo caused by the aforementioned processes. Or put differently, if the kinematic evolution of galaxies is indeed tied to its halo, the cluster environment affecting the halo could eventually also disrupt the kinematics of galaxies at its centre. This question now lies at the heart of the present study.
    
    Previous numerical studies have shown how tidal interactions can disrupt infalling haloes in different environments. The seminal work of \citet{Hayashi03} studied how tides influence substructure in dark matter-only simulations. Their analysis showed that, although tides preferentially strip the outer regions of haloes, they also decrease the halo's central density after each pericentric passage. Subsequent studies improved the tidal disruption estimation by including a stellar component to their analysis  \citep[e.g.][]{Bullock05,Penarrubia08b}. The stellar component was found to be exceptionally resilient to tides, preserving its density profile shape even after losing a considerable amount of stars. Nevertheless, the stellar dynamics of such calculations were modelled by analytic profiles, which simplify processes such as the mass loss from dynamical friction, the halo phase-space evolution after merger interactions, or the stellar mass fractions determined by star formation. More recent studies overcame these limitation by introducing a stellar component modelled by full-physics hydrodynamics \citep[e.g.][]{Smith16,Lokas20,Errani20,Mazzarini20}. However, these studies primarily focus on mass-loss processes, hence the question of how environmental effects influence the kinematic properties of the stellar component of infalling haloes remains to be addressed.

    We approach these issues by analysing simulations from \textsc{The Three Hundred} project\footnote{\url{https://the300-project.org}}, i.e. a sample of over 300 galaxy clusters simulated with full-physics hydrodynamics \citep{Cui18}. These simulated clusters have been used for different studies, for example, environmental effects \citep{Wang18}, cluster profiles \citep{Mostoghiu19,Li20,Baxter21}, backsplash galaxies \citep{Arthur19,Haggar20,Knebe20}, cluster dynamical state \citep{Capalbo20, DeLuca20}, filament structures \citep{Kuchner20,Rost21,Kuchner21} and gravitational lensing \citep{VegaFerrero21}. In this work we extend the previous stellar angular momentum analysis to this set of massive simulated galaxy clusters to study the influence of the cluster environment on the stellar kinematics of infalling galaxies.
    
    This paper is organised as follows. In \Sec{sec:data} we present the data used for the analysis. \Sec{sec:data-subsec:simulations} briefly describes the simulations. In \Sec{sec:data-subsec:sample} to \Sec{sec:data-subsec:stellarcomponent} we define how we selected the sample of galaxies used for the analysis and we present the classification of our objects. \Sec{sec:results} describes our results: in \Sec{sec:results-subsec:kappa} we focus on the kinematic evolution of the sample, and in \Sec{sec:results-subsec:jM} we study their angular momentum-stellar mass relation. Finally, we conclude the study in \Sec{sec:conclusions}.


\section{The Data} \label{sec:data}

    \subsection{`The Three Hundred' Central Galaxy Clusters} \label{sec:data-subsec:simulations}
    \paragraph*{The Simulations}The simulated clusters in \textsc{The Three Hundred} dataset were created by extracting 324 spherical regions of $15 h^{-1}$ Mpc radius centred on each of the most massive haloes identified at $z=0$ within the dark-matter-only MDPL2 simulation \citep{Klypin16}\footnote{The MultiDark simulations -- incl. the MDPL2 used here -- are publicly available at \url{https://www.cosmosim.org}}. MDPL2 was simulated with a \textit{Planck} 2015 cosmology \citep{Planck15}, with $\Omega_{\rm M} = 0.307$, $\Omega_{\rm b} = 0.048$, $\Omega_\Lambda = 0.693$, $h=0.678$, $\sigma_8 = 0.823$, and $n_s=0.96$ and consists of a box of $1 h^{-1}$ Gpc side-length which contains $3840^3$ dark matter particles each of mass $1.5 \times 10^9 h^{-1}$ M$_{\odot}$. In order to model the galaxy clusters with all the relevant baryonic physics, those $15 h^{-1}$ Mpc regions were traced back to the initial conditions and there populated with gas particles by leading to a mass resolution of $m_{\rm DM}=1.27\times 10^{9} h^{-1}$$\Msun$ and $m_{\rm gas}=2.36\times 10^{8} h^{-1}$$\Msun$, respectively. Outside the re-simulated region, to reduce the computational cost of the original MDPL2 simulation, dark matter particles are degraded with lower mass resolution particles to maintain the same large scale tidal field. Using a Plummer equivalent softening of $6.5h^{-1}$ kpc for both the dark matter and baryonic component, the new initial conditions were now moved forward in time using \code{GADGET-X}  \citep{Beck16}.  \code{GADGET-X} is  a modified version of \code{GADGET3}  with a modern Smooth Particle Hydrodynamics (SPH) scheme which improves the treatment of gas particles \citep{Beck16, Sembolini16b}. Results of simulations of galaxy clusters based on \code{GADGET-X} have been presented in several previous papers \citep[e.g.][]{Rasia15,Planelles17} and in the \textit{nIFTy cluster comparison} project \citep{Sembolini16a,Elahi16,Cui16,Arthur17}. A total of 129 snapshots are saved from $z=16.98$ to $z=0$.
    
    \paragraph*{The Halo Finding}The halo analysis was done using the \code{AHF}\footnote{\url{http://popia.ft.uam.es/AHF}} halo finder \citep{Gill04a,Knollmann09}. \code{AHF} locates local overdensities in an adaptively smoothed density field as potential halo centres and automatically identifies haloes and substructure (subhaloes, subsubhaloes, etc.). The radius of a halo $R_{200}$ and the corresponding enclosed mass $M_{200}$ are calculated as the radius $r$ at which the cumulative density $\rho(<r)=M(<r)/(4\pi r^{3}/3)$ drops below $200\rho_{\rm crit}(z)$, where $\rho_{\rm crit}(z)$ is the critical density of the Universe at a given redshift $z$. 
    
    \paragraph*{The Merger Trees}The progenitors of the haloes are tracked across the snapshots with \code{MergerTree}, a tool that comes with \code{AHF}. Each halo identified at redshift $z = 0$ is tracked backwards in time, identifying as the main progenitor at some previous redshift the halo that maximises the merit function $\mathcal{M} =N_{A\cap B}^2/(N_{A} N_{B})$, where $N_A$ and $N_B$ are the number of particles in haloes $H_A$ and $H_B$, respectively, and $N_{A\cap B}$ is the number of particles that are in both $H_A$ and $H_B$. The code further has the ability to skip snapshots, i.e. progenitors of haloes that are not found in the previous snapshot are still searched for in earlier snapshots, recovering an otherwise truncated branch of the merger tree \citep[see][]{Wang16}. However, to reduce errors during the tracking of the stellar kinematic evolution of our objects we consider an object `lost' if it cannot be found for 5 consecutive snapshots.\\
    
    In summary, we have 324 numerically modelled central galaxy clusters and all the haloes orbiting in and about them out to a distance of $15 h^{-1}$ Mpc available for our analysis. The details of this full data set are presented in \citet{Cui18}. Here we are though applying a few selection criteria to both the central clusters and the field haloes and subhaloes in the regions, to be described now.

    \subsection{Cluster Selection} \label{sec:data-subsec:sample}
    As we aim at tracing back all the objects in and about the central galaxy cluster, our analysis requires that we can always define a main progenitor for each of the 324 central haloes. As shown in \citet{Behroozi15}, major mergers during the formation of those central objects pose a challenge to this. Following the discussion in \citet{Haggar20}, we therefore identified and removed central cluster haloes whose main progenitor's position changes by more than half their radius $R_{200}(z)$ between two consecutive snapshots $[z,z+\Delta z]$. We further require the main branch to at least extend to redshift $z=2$. This reduces the number of regions entering the analysis from 324 down to 236. This is approximately 8 times more cluster regions than used in other state-of-the-art cluster simulation studies \citep[e.g.][]{ Bahe17,Barnes17}.

    \subsection{Halo Selection} \label{sec:data-subsec:haloselection}
    Each of our 236 selected central galaxy clusters is surrounded by a multitude of haloes out to the  $15 h^{-1}$ Mpc edge of the region that was modelled including all the relevant baryonic physics. But as we are interested in studying the angular momentum of the stellar component of these haloes, we are limiting our analysis to haloes with at least $M_{\rm {star}} > 5\times10^{10}\hMsun$. This corresponds to at least $1000$ star particles. Further, only those haloes that lie within $3R_{200}$ of the central galaxy cluster at redshift $z=0$ are traced backwards in time. For these haloes we then define their infall redshift $z_{\rm{inf}}$ as the redshift at which the halo crosses central galaxy cluster's $2R_{200}(z)$ for the first time: recent numerical studies show that around $\sim 1.5R_{200}-2R_{200}$, haloes experience a sharp cut-off in their gas content which could indicate the presence of an accretion shock \citep{Power18, Arthur19, Baxter21} and hence we decided to use $2R_{200}$ (instead of $R_{\rm 200}$) as our reference crossing radius. Haloes that cannot be assigned an infall redshift will be removed from the analysis.\\

    \subsection{Stellar Component} \label{sec:data-subsec:stellarcomponent}
    To select the stellar component of our objects -- that one might identify with their central galaxy -- we opted for a spherical region enclosing 15 per cent of the haloes' physical radius \citep[e.g.][]{Bailin05}. However, as tidal interactions with the central galaxy cluster impact on the size of subhaloes \citep[][]{Muldrew11,Onions12} we used the radius as found right before crossing $2R_{\rm 200}$ of the central galaxy cluster. This aperture is now interpreted as the `size of the galaxy' and kept fixed in physical coordinates across the snapshots. We performed a series of tests using different criteria (e.g. an aperture not depending on the halo's physical radius but rather fixing it to 30kpc) that showed that even though haloes suffer an overall mass loss while orbiting within their host halo, the central stellar mass in the aperture is mostly unaffected. This agrees with other simulations, which show that stellar stripping is rare, and happens only after the dark matter has suffered significant stripping \citep[$> 80$ per cent, e..g][]{Smith16,Bahe19}. It is important to point out again that even though the stellar mass of our galaxy does not change (something we also quantify below), we cannot rule out any back-reaction of the stars to the change in potential caused by tidal stripping of the halo. Finally, to take into account the spatial resolution of our simulations we restrict our objects to have a `galaxy size' of at least $2\epsilon = 13\hkpc \sim 20$ kpc. We remark that this additional condition removes only the smallest galaxies in the sample, which is less than 5 per cent of the objects selected so far. Moreover, practically all of the star particles reside within our aperture and hence we are not considerably reducing the number of star particles by cutting out the `galaxy' as defined here. The remaining 6509 objects, with halo masses from $9.2\times10^{10}\hMsun$ to $5.3\times10^{14}\hMsun$, constitute the analysis sample, built by combining all the objects that satisfy the aforementioned criteria from the 236 cluster regions considered here.


\section{Results} \label{sec:results}
    In what follows we are comparing the properties of our selected haloes orbiting in and about the central galaxy clusters at infall redshift $z_{\rm inf}$ (i.e. when crossing $2R_{\rm 200}$) and present day time $z=0$.

    \subsection{Mass Evolution}\label{sec:results-subsec:mass}
    \begin{figure}
    \includegraphics[width=\columnwidth]{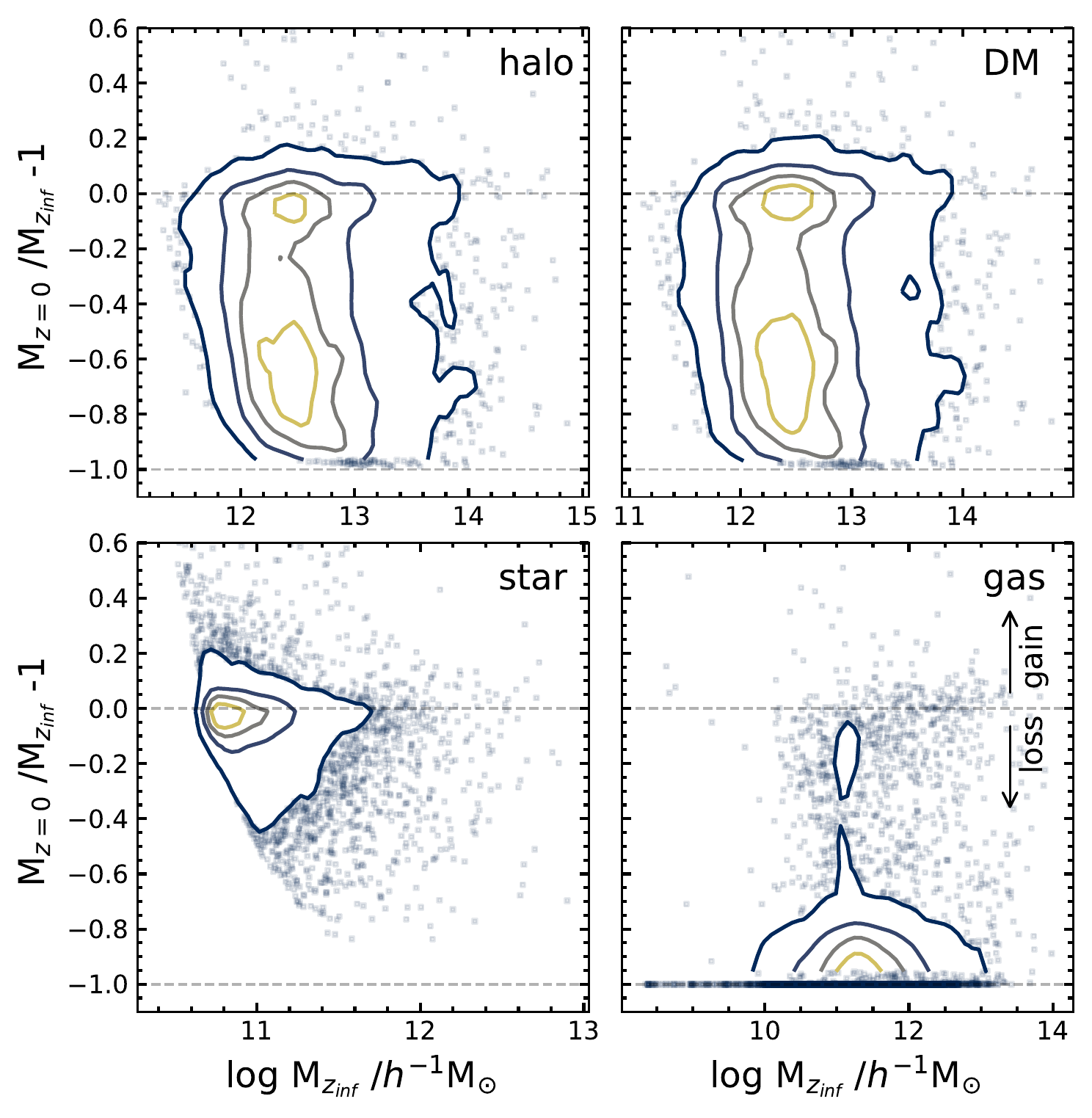}
    \caption{Evolution of the different mass components of the haloes in the sample from their infall redshift $z_{\rm{inf}}$ to their $z=0$ values. Contour levels show the 75, 50, 25 and 5 per cent of the maximum counts of the distributions.  The horizontal lines at 0 and -1 correspond to no change and maximum mass loss, respectively. The position in the plane of objects beyond the 5 per cent contour are shown as square markers.}
    \label{fig:deltam}
    \end{figure}
    
    To understand the processes that shape the haloes after their first infall we show in \Fig{fig:deltam} the fractional change in mass since their infall redshift and as a function of infall mass. Contour levels show 75, 50, 25, and 5 per cent of the maximum counts of the distributions. The horizontal lines at 0 and -1 correspond to no change and maximum mass loss, respectively. Objects beyond the 5 per cent contour are shown as square markers. As expected, haloes crossing the denser regions of the central galaxy cluster lose a considerable amount of mass  \citep[e.g.][]{Klimentowski10}. Haloes that only entered the central halo once and did not leave, in general, conserve more of their initial mass than the ones that experienced multiple infalls into the denser regions of the galaxy cluster. In terms of the stellar mass in the haloes, the loss is significantly lower due to the fact that most of the stellar component resides in the central region of the halo mainly shielded from tidal forces. We nevertheless observe an increase in stellar mass loss at the 5 per cent level as we move to the massive end of the stellar mass distribution. Such haloes present substructure residing in the halo component (i.e. outside the central aperture) and thus they are prone to suffering mass loss from processes that strip the halo at its outskirts. The gas component, on the other hand, is mostly gone regardless of their mass, as gas is affected by an entirely different set of processes as they fall into the central halo, i.e. ram-pressure stripping \citep{Arthur19, Mostoghiu21}.  Studying the radial distribution of the gas inside infalling haloes (not shown here), we find that haloes that never reached the denser regions of the cluster halo are still considerably stripped of their gas component, in agreement with \citet{Power18, Arthur19, Mostoghiu21}.\\

    In summary, \Fig{fig:deltam} clearly shows that the stellar component of our haloes is the least affected by the environment of the central galaxy cluster. The question nevertheless remains if we will find changes in the kinematical properties. But the haloes of our `galaxies' (as defined by the star particles in the aperture, cf. \Sec{sec:data-subsec:stellarcomponent}) certainly undergo some changes as manifested by the mass loss. It is therefore not yet clear that even though the galactic stellar mass more or less remains constant that there will be no reaction of its internal dynamics to the varying influence of the central galaxy cluster.

    \subsection{Fraction of co-rotational energy}\label{sec:results-subsec:kappa}
    We study the influence of the central galaxy cluster on the internal dynamics of our galaxy sample after their first infall by following the evolution of their specific stellar angular momentum $j_{*}$.
    
    The galaxies in our sample can be classified by the fraction of stellar kinetic energy that is invested in co-rotation $\kappa_{\rm{corot}}$, as presented in  \citet{Sales10,Correa17}:
    \begin{equation}\label{eq:kappa_def}
        \kappa_{\rm{corot}} = \frac{E_{\rm{corot}}}{E_{\rm{kin}}}\textrm{   ,   }
        E_{\rm{corot}}=\sum_{j_{z,i}>0}{ \frac{1}{2}m_{i}\left(\frac{j_{z,i}}{r_{2d,i}}\right)^2}\textrm{   ,   }
    \end{equation}
    where $E_{\rm{kin}}$ is the kinetic energy of the star particles in the aperture, $E_{\rm{corot}}$ is the rotational energy of the corotating star particles contributing to the rotation of the galaxy, $m_{i}$ the mass of the star particle;
    \begin{equation}
    \begin{split}
        j_{z,i}   &=\bm{j}_{i}\cdot\bm{\hat{J}_{\rm{tot}}}\textrm{ ,} \\
        \bm{j}_{i} &=(\bm{r}_i-\bm{r}_{\rm{halo}})\times(\bm{v}_i-\bm{v}_{\rm{halo}})\textrm{ , and } \\
        \bm{\hat{J}}_{\rm{tot}} &= \frac{1}{J_{\rm{tot}}} \frac{\sum{m_{i}\bm{j}_i}}{\sum{m_{i}}}\textrm{   ,   }
    \end{split}
    \end{equation}
    the specific angular momentum along the direction of the total angular momentum, the specific angular momentum of a star particle in the rest frame of its halo, and the direction of the total specific angular momentum of the galaxy, respectively; and $r_{2d,i}=(\lvert\bm{r}_i-\bm{r}_{\rm{halo}}\rvert^2 -((\bm{r}_i-\bm{r}_{\rm{halo}}) \cdot \bm{\hat{J}}_{\rm{tot}})^2)^{1/2}$ the cylindrical radius of star particles. As we exclusively use the definition based on co-rotating star particles, we drop the subscript from the parameter $\kappa$ from now on. 

    \subsubsection{Changes in $\kappa$ since infall}\label{sec:results-subsec:kappa-subsubsec:kappainf}
    
    \begin{figure}
    \includegraphics[width=\columnwidth]{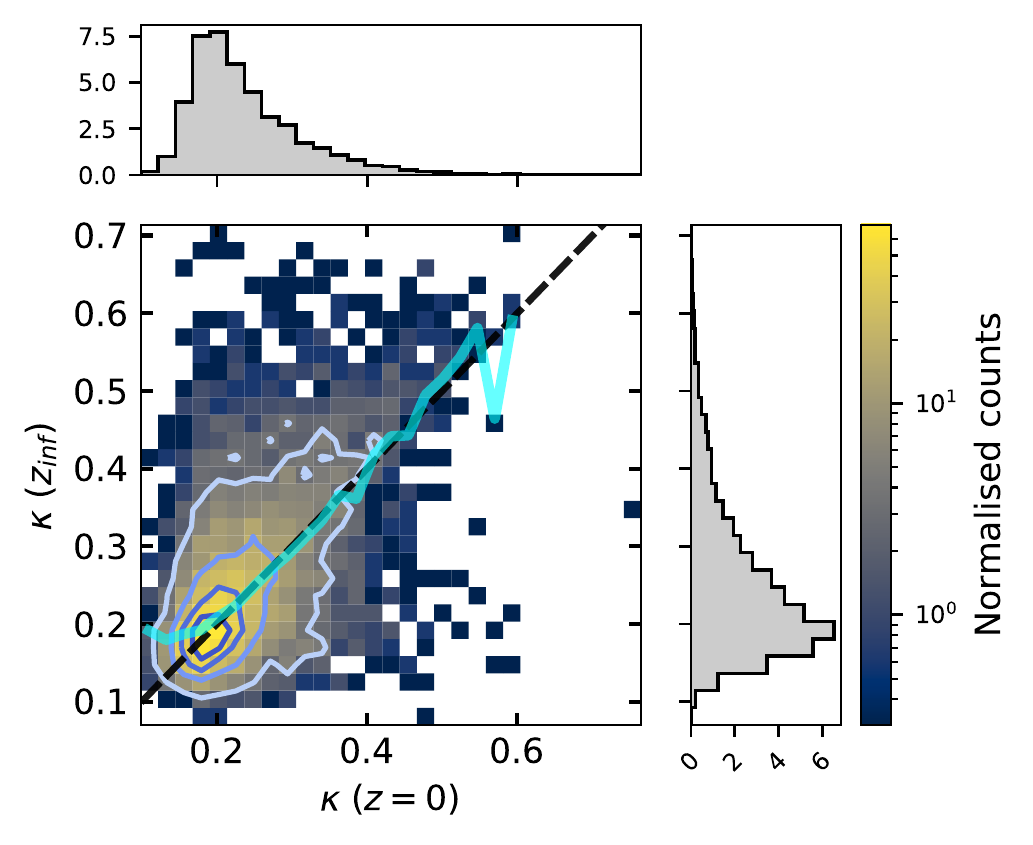}
    \caption{Fraction of ordered rotation $\kappa$ at $z=0$ and at their infall redshift $z_{\rm{inf}}$. The diagonal dashed line shows the $1:1$ relation.  Median values in bins of $\kappa (z=0)$ for the sample are represented by the solid cyan line. Bins are coloured by their corresponding normalised counts. The contour lines show where 75, 50, 25, and 5 per cent of the maximum counts lie.}
    \label{fig:kappaCorotAp_compare}
    \end{figure}
    
    We compared the fraction of ordered rotation at two different times in the evolution of the galaxies to find if the $z=0$ values are the result of the influence of the central galaxy cluster during their infall. In \Fig{fig:kappaCorotAp_compare} we show the $\kappa$ values of the sample at $z=0$ compared to their values at infall $z_{\rm{inf}}$, before experiencing the central cluster's influence. The black dashed diagonal line shows the $1:1$ relation. The median values in bins of $\kappa$ at $z=0$ are represented by the solid cyan line. Bins are coloured by their corresponding number of counts and contours show where 75, 50, 25, and 5 per cent of the maximum counts lie. 
    
    We find that for most of the haloes in the sample the fraction of stellar ordered rotation is unaffected by their infall (Pearson coefficient of 0.56), with a median value at $z=0$ and at $z_{\rm{inf}}$ of $\kappa\sim 0.22$. However, beyond the 5 per cent contour we identify galaxies which experienced a considerable change in $\kappa$ since their infall, e.g. $\Delta\kappa \sim \pm0.2-0.4$.
    
    \begin{figure}
    \includegraphics[width=\columnwidth]{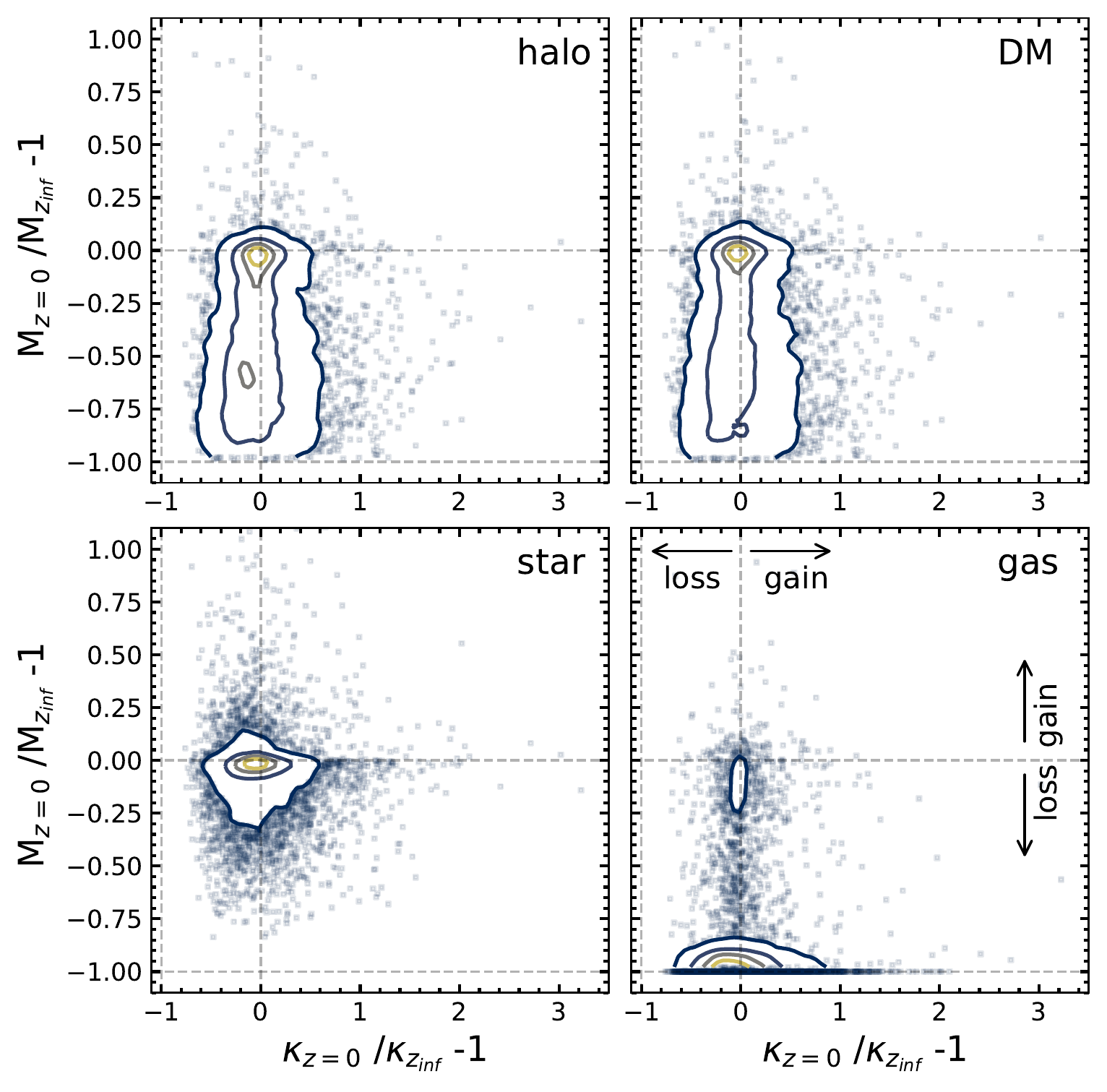}
    \caption{Evolution of the different mass components of the haloes in the sample and their fraction of ordered rotation $\kappa$ from their infall redshift $z_{\rm{inf}}$ to their $z=0$ values. Contour levels show the 75, 50, 25 and 5 per cent of the maximum counts of the distributions.  The horizontal lines at 0 and -1 correspond to no change and maximum mass loss, respectively. The position in the plane of objects beyond the 5 per cent contour are shown as square markers.}
    \label{fig:deltamdeltakappa}
    \end{figure}
    
    In \Sec{sec:results-subsec:mass} we found that the haloes in the sample are stripped of their material during their fall into the into the denser regions of the central galaxy cluster. Nevertheless, mass-loss does not appear to considerably affect the rotational properties of their galaxies, as seen from their $\kappa$ parameter. We quantify this change in \Fig{fig:deltamdeltakappa}, in which we show the same evolution of the mass component of the haloes in the sample presented in \Fig{fig:deltam} but this time as a function of the evolution in their fraction of ordered rotation. 
    
    For the total halo and DM mass components of the haloes in the sample we identify objects that lost up to 90 per cent of their mass at infall, yet we do not find a correlation between the amount of mass-loss and their fraction of ordered rotation evolution trend. At the 5 per cent contour, regardless of the amount of stripped mass, their fraction of stellar ordered rotation changed within $\sim\pm60$ per cent since their infall. As we discussed in the previous section, the stellar component is less affected by mass-loss processes. Just so, we see that at the 5 per cent contour the objects experienced a change in stellar mass within $\sim\pm30$ per cent and that their change in $\kappa$ tends towards zero as their infall mass is disrupted. This is likely due to the fact that such stellar mass gain/loss tends to happen in the outer region of haloes, outside the apertures used to define the central galaxy and calculate $\kappa$. As for the gas component, even when haloes have been mostly stripped of their gas by the time they reach $z=0$, the stellar fraction of ordered rotation does not correlate with such mass-loss. Our galaxies can fall into galaxy cluster regions, losing most (if not all) of their gas during their infall, and still retain the stellar kinematic properties they had prior their infall. Similar results have been reported in \citet{Cortese19}: satellite galaxies go through significant changes in their specific star formation rate, but those are not necessarily accompanied by changes in their stellar spin parameter.

    Overall, we find no correlation between the amount of stripped mass and the change in their $\kappa$ parameter. Thus, we conclude that the fraction of ordered rotation in our simulations is hardly affected by the (potentially violent) stripping processes that disrupt the dark matter halo.

    \subsubsection{Temporal evolution of high and low $\kappa$ galaxies}\label{sec:results-subsec:kappa-subsubsec:kappaovertime}
    In \Fig{fig:kappaCorotAp_compare} we identified objects beyond the 5 per cent contour with a considerable change in their fraction of stellar co-rotational since their infall. To study which processes are responsible for such changes we selected from the $\kappa$ distribution at $z=0$ the galaxies within 5 per cent of the highest and lowest fractions of ordered rotation, i.e. the $\kappa > 95$-th percentile, the high $\kappa$ sub-sample; and $\kappa < 5$-th percentile, the low $\kappa$ sub-sample of the distribution. This corresponds to a high $\kappa$ threshold of 0.38 and a low $\kappa$ threshold of 0.15. Note that, our choice for the threshold values is purely motivated by kinematics: we simply aim to understand the origin of the substantial changes in the $\kappa$ distribution of a (relatively) small fraction of the total sample.  Adopting other threshold values \citep[which also take into account other non-kinematic properties such as star formation or colours, e.g. $\kappa = 0.4$ from][]{Correa17} for the stellar kinematic classification, only 4 per cent of our sample would be classified (in terms of their kinematics) as fast-rotating galaxies. 
    
\begin{figure}
    \includegraphics[width=\columnwidth]{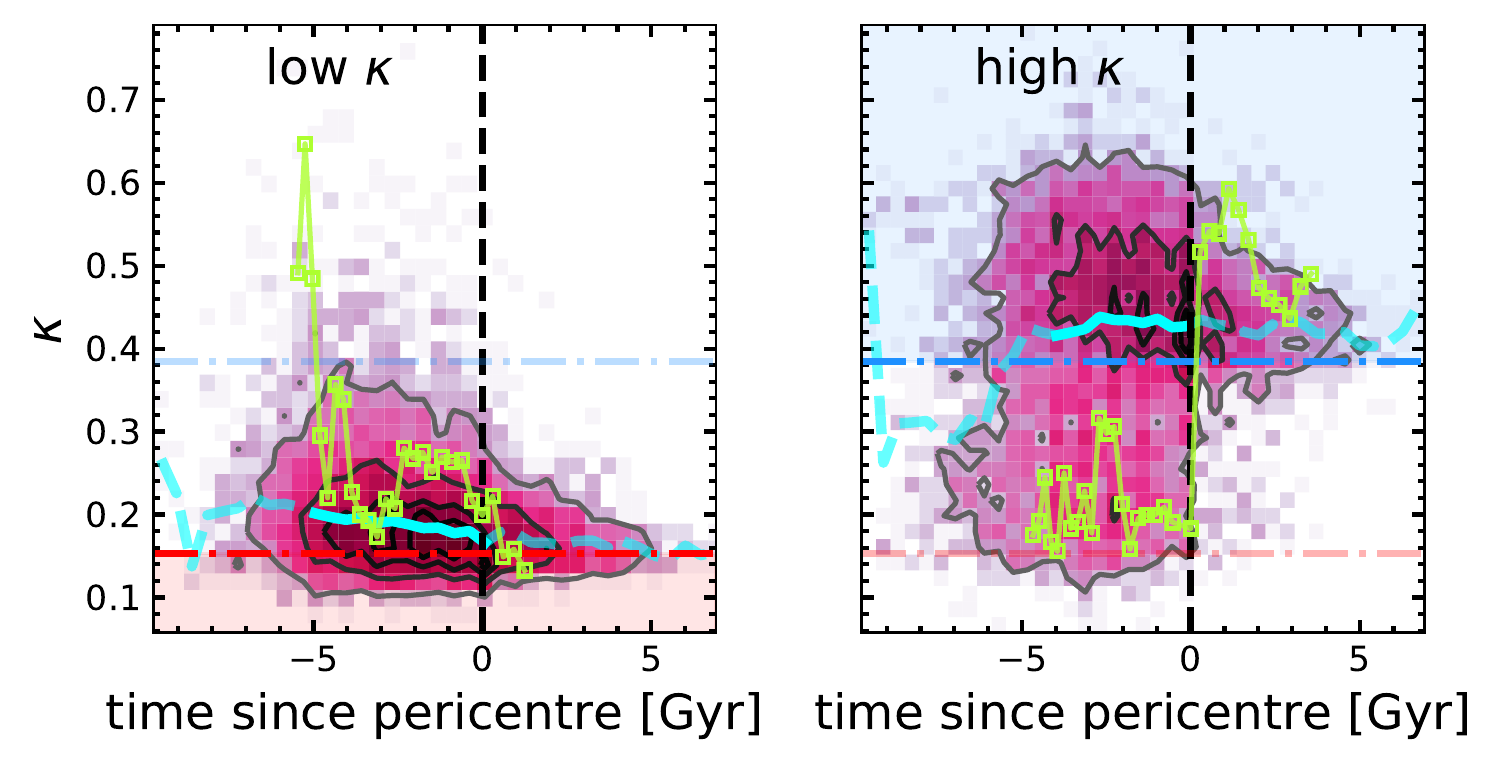}
    \caption{Evolution of the fraction of co-rotational energy as a function of the time since the pericentre passage for low (left) and high (right) $\kappa$ sub-sample as classified by the $5-$th and $95-$th percentiles of the $z=0$ $\kappa$ distribution in \Fig{fig:kappaCorotAp_compare}. Contours show 75, 50, 25, and 5 per cent of the maximum counts of the 2D distribution. The cyan solid line shows the median value at each time bin. For bins in which less than 50 per cent of the sample entered the median calculation, we instead use a dashed line. The red and blue horizontal dash-dotted line shows the $\kappa$ threshold values for the low and high $\kappa$ sub-samples, respectively. The pericentre is marked with a vertical dashed line. In each panel we additionally show the evolution of a galaxy from the corresponding sub-sample using green square markers. The shaded red and blue regions show the allowed $\kappa$ values corresponding to each sub-sample.}
    \label{fig:kappa_peritime}
    \end{figure}

    Recent numerical results show that galaxies which continue to accrete gas and form stars are very efficient at spinning up \citep{Lagos17}. However, this is unlikely to be the cause here as gas accretion is expected to be hampered in clusters. \citet{Lagos18a} found that galaxies can be spin up or down by mergers depending on their orbital orientation and gas content \citep[see][for similar results]{Schulze18, Lagos18b}. To investigate the origin of the change in the fraction of co-rotation after infall found in \Fig{fig:kappaCorotAp_compare}, in \Fig{fig:kappa_peritime} we study the $\kappa$ evolution of the low and high $\kappa$ sub-samples found at $z=0$ as a function of the time since their pericentre passage, defined as the closest approach of an infalling halo to the central galaxy cluster \footnote{Our definition of `pericentre' does not necessarily imply they are within $R_{200}$ of the central galaxy cluster.}. Contours show 75, 50, 25, and 5 per cent of the maximum counts in the distribution. We show the median value at each time bin with a cyan solid line, and when the number of haloes in a bin is less than 50 per cent of the maximum count we use a dashed line instead. The low and high $\kappa$ threshold values are represented by the red and blue horizontal dash-dotted lines, respectively, and the pericentre time is marked with a vertical dashed line. In each panel we additionally show the $\kappa$ evolution of a galaxy from the respective sub-sample with green square markers, and we marked the allowed $\kappa$ values of each sub-sample with corresponding shaded regions.  
    
    As already anticipated by the $\kappa$ evolution in \Fig{fig:kappaCorotAp_compare}, most of the galaxies in each sub-sample retain their fraction of co-rotational energy $\kappa$ after their infall. However, over the course of 5 Gyrs after the pericentre passage we observe a decrease of $\sim0.2$ in $\kappa$ in both sub-samples ($\sim0.04/$Gyr) at the 5 per cent contour. Moreover, the $\kappa$ evolution of the low $\kappa$ sample galaxy shows that it had high stellar co-rotational energy at some point, but lost it after $\sim6$ Gyr. Such slow decrease can be attributed to the two-body heating of the stellar component: as galaxies infall into the central galaxy cluster, the background (more massive) dark matter particles residing inside the cluster that fly by these galaxies tend to increase the stellar mean interparticle distance due to the softening scale used for the stellar particles, heating up the stellar component and effectively puffing up the stellar distribution while spinning them down. As these galaxies fall further into the cluster, the density of dark matter particles increases and consequently the amount of fly-bys is boosted. Along with this slow decrease, we also identify processes which instead can spin-up galaxies on a faster scale. But these changes -- as observed for the high $\kappa$ sample galaxy -- are happening close to pericentre passage and are investigated in more detail now.

    \subsubsection{Spinning-up galaxies with low $\kappa$}\label{sec:results-subsec:kappa-subsubsec:spinup}
    As two-body heating slows down the stellar kinematics of every infalling galaxy in equal measure, i.e. $\kappa$ decreases at the same rate for both sub-samples, we focus on processes which are able to spin-up galaxies during their infall. To isolate the galaxies which cross the threshold $\kappa$ value to become high $\kappa$ galaxies according to our classification, from the ones that conserve their classification up to $z=0$, i.e. the ones that remained within the high $\kappa$ sub-sample after their infall into the cluster environment, we select galaxies from the high $\kappa$ sub-sample which experienced a rapid $\kappa$ increase (i.e. $\Delta\kappa > 0.2$) within 1 Gyr since their pericentre passage. These galaxies form only 12 per cent of the high $\kappa$ galaxies at $z=0$ (39 objects, 0.6 per cent of the total number of objects in the sample). Within this new sub-sample, we identify galaxies for which a sudden change in their star particle count was accompanied by a sudden spin-up of their specific angular momentum (and consequently an increase in their $\kappa$ parameter) near their pericentre, and galaxies for which the number of stars close to the pericentre does not seem to play a crucial role in their $\kappa$ evolution.
    
    Following the star particle distribution of these galaxies, we find that the fast increase in $\kappa$ and in the number of stars is a consequence of two processes: mergers with other haloes, and fly-bys of stellar remnants within the central aperture used to define the galaxies residing in each halo. In agreement with \citet{Lagos18a}, we find that the co-rotating (counter-rotating) infalling stars from mergers are able to spin-up (spin-down) our galaxies. On the other hand, acting on shorter timescales and without a significant stellar gain, we find that stellar remnants are able to temporarily disrupt the spin of our objects. These haloes are the residual cores of infalling haloes, where the original dark matter component of the halo has become lost and subsumed by the main halo. Considering that the stellar component of the remnant haloes is smaller than the central galaxy extension of the haloes in the sample ($\sim 1/3$ of the aperture size), and that the amount of remnant stellar cores increases close to the centre of the galaxy cluster, these objects contribute transiently to the co-rotational energy in the aperture and, as such, $\kappa$ increases near the pericentre of our objects.\\

    \subsubsection{Summary}\label{sec:results-subsec:kappa-subsubsec:summary}
    We conclude that for most of the galaxies in our sample entering massive galaxy clusters, their kinematic properties (as captured by the fraction of ordered rotation $\kappa$) do not change in a significant manner -- even though we have seen in \Sec{sec:results-subsec:mass} that their haloes undergo substantial changes. In general, low $\kappa$ galaxies at $z=0$ had low fractions of ordered rotation before entering the cluster halo, and high $\kappa$ galaxies at $z=0$ had such high fractions prior to entering the cluster environment. The two-body heating of the stellar component of infalling haloes induced by the more massive dark matter particles in the cluster environment affects equally both samples, slowly reducing their stellar specific angular momentum as they orbit the cluster region. As such, we find galaxies that at their infall time had high fractions of stellar co-rotational energy ($\kappa \gtrapprox 0.4$) that ended up as low $\kappa$ galaxies by the time they reached $z=0$, i.e. with a $\kappa<5-$th percentile of the $\kappa$ sample distribution at $z=0$. On the other hand, from the high $\kappa$ sample at $z=0$, i.e. the galaxies with $\kappa>95-$th percentile, we identify $\sim12$ per cent of them which previously had considerably lower fractions. We find that these two processes acting on different timescales, i.e. mergers with other infalling haloes and stellar remnants transiting the aperture used for defining the galaxies of the haloes in the sample, are able to spin-up the galaxies.

    
    \subsection{Stellar angular momentum and stellar mass relation}\label{sec:results-subsec:jM}
    \begin{figure}
    \includegraphics[width=0.9\columnwidth]{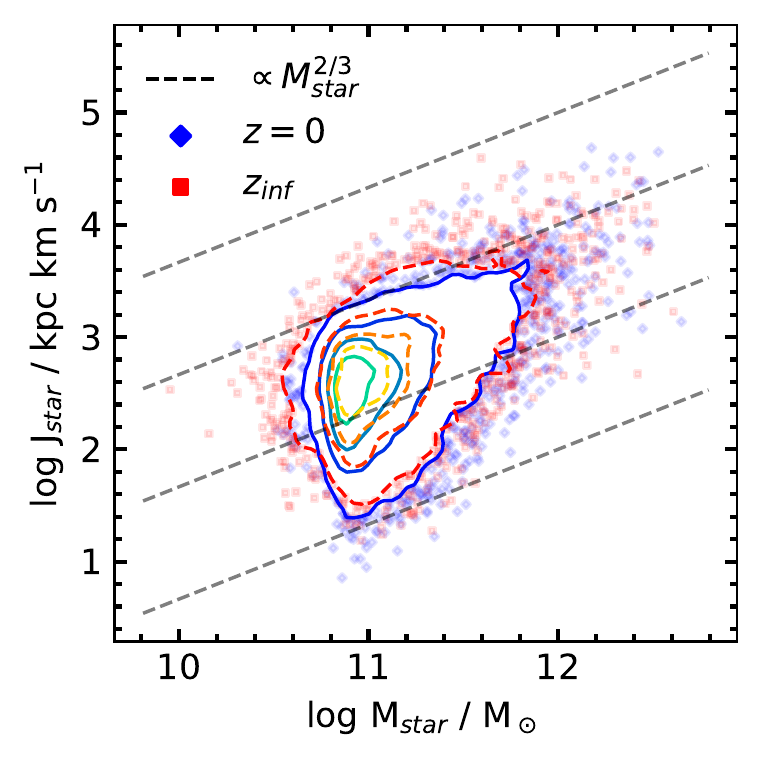}
    \caption{Specific angular momentum and stellar mass relation of the sample of galaxies at $z=0$ and at their infall redshift $z_{\rm{inf}}$. The diagonal dashed lines show the $M_{\rm{star}}^{2/3}$ relation for zero-points $-3$ (top line) to $-6$ (bottom line). The contours show 75, 50, 25 and 5 per cent of the sample. The position in the plane of objects beyond the 5 per cent contour are shown as filled markers.}
    \label{fig:jstarmstar_zcompare}
    \end{figure}
    
    In the previous section we identified processes which are able to change specific angular momentum of infalling galaxies. Nevertheless, most of the objects remain unaffected by such interactions. In this section we study the influence of the central galaxy cluster on the internal dynamics of our galaxies by following their evolution in the specific angular momentum $j_{\rm{star}}$ -stellar mass M$_{\rm{star}}$ plane, an extensively studied fundamental kinematic property of galaxies which can be described a power-law of the form $j_{\rm{star}}\propto$ M$_{\rm{star}}^{\alpha}$, where $\alpha \approx 0.7$ \citep[e.g.][]{Fall83, Romanowsky12, Obreschkow14, Teklu15, Fall18}.
    
    As some of our galaxies experienced great changes in their fraction of ordered rotation during their evolution since their infall, we aim to find if these changes are also visible in the specific stellar angular momentum-stellar mass plane. We show their position in the $j-M$ plane after their infall redshift in \Fig{fig:jstarmstar_zcompare}. The contours represent 75, 50, 25, and 5 per cent of the sample. The position in the plane of galaxies beyond the 5 per cent contour are shown as filled markers. The dashed lines indicate the $j_{\rm{star}} = \lambda M_{\rm{star}}^{2/3}$ relation for different zero-points ($\log$ $\lambda$ = $-3$ to $-6$), as presented in \citet{Teklu15}. Note that, overall, the sample in this analysis does not include galaxies with stellar specific angular momenta typical of disk galaxies in the local Universe.
    
    Once again, we find that the distribution is mostly unchanged after entering the cluster halo. However, galaxies positioned furthest from the average position in the $j-M$ plane at their infall redshift move towards lower specific angular momentum and stellar mass values after infalling, as seen in the 5 per cent contours and points below that level. 
    
    These results show that the internal dynamics of our infalling galaxies into massive galaxy clusters, as quantified by either the fraction of ordered rotation $\kappa$ or the evolution of their location in the specific angular momentum-stellar mass plane, are mostly unaffected by the processes in such environments even after their halos have been considerably stripped of their mass, and consequently, disrupted the halo potential in which these galaxies reside.


\section{Conclusions} \label{sec:conclusions}
    We analysed the central stellar component of haloes from \textsc{The Three Hundred} project, a suite of 324 galaxy cluster regions simulated with full-physics hydrodynamics, to study of the influence of galaxy clusters on the internal dynamics of infalling galaxies. We selected objects with M$_{\rm{star}}>5\times10^{10}$ $\hMsun$ within $3R_{200}$ from the galaxy cluster halo at $z=0$. After applying different selection criteria that ensure the correct tracking of the stellar component of each halo we obtain a sample of 6509 galaxies from 236 cluster regions, i.e. approximately 8 times more cluster regions than used in other state-of-the-art cluster simulation studies \citep[e.g.][]{Bahe17,Barnes17}. Here we summarise our main results:
    
    \begin{itemize}
    
        \item Using the fraction of stellar co-rotational energy $\kappa$ to track the internal dynamics of the infalling galaxies in our sample we find that it remains mostly unchanged for most of the galaxies after they enter the galaxy cluster environment, despite the (potentially violent) stripping processes that disrupt their haloes.

        \item We studied the time evolution of the $\kappa$ parameter. Even though most galaxies do not experience great $\kappa$ changes during their evolution, we identify an overall slow decrease in the fraction of ordered rotation close to their pericentre passage ($\sim0.04/$Gyr) which we attribute to numerical effects. Along with this effect we found that $\sim12$ per cent of the galaxies with high $\kappa$ at $z=0$ (i.e. $\kappa>95-$th percentile of the $z=0$ distribution), an already small population in our sample, had considerably lower fractions of co-rotational energy in the past and experienced a fast boost in their $\kappa$ parameter within the past few Gyrs ($\sim0.2/$Gyr) due to two different processes: stellar mergers of infalling haloes \citep[in agreement with ][]{Lagos18b,Lagos18a}, and transient encounters with the stellar remnants of haloes that have lost their dark matter component during their own passage through the cluster, entering within the aperture we used for defining the properties of our galaxies. 

        \item We found that similar to the $\kappa$ evolution of our galaxies, the specific angular momentum-stellar mass relation for the galaxies in the sample showed no substantial change in their location on the $j-M$ plane from their infall redshift until $z=0$.
    \end{itemize}
    
    Previous numerical studies \citep[e.g.][]{Smith16,Lokas20} have shown that infalling and orbiting haloes within galaxy clusters are disrupted by a series of processes which dominate in such environments, such as ram-pressure stripping \citep[e.g.][]{Bahe15,Arthur19, Mostoghiu21}, mergers \citep[e.g.][]{Behroozi14,Lagos18a} or tidal torques \citep[e.g.][]{Park07}. Nevertheless, these numerical studies primarily analyse mass-loss processes in infalling haloes. Considering that cluster environments disrupt the different mass components of infalling haloes, such environments could also potentially disrupt their stellar kinematics. 
    
    In this work we investigate such question by studying the internal central stellar dynamics of infalling objects towards the numerically simulated massive galaxy clusters in our sample. We found that, roughly, their $z=0$ classification holds even from before the time they entered the cluster environment. Moreover, we find that in terms of their position in the fundamental specific stellar angular momentum-stellar mass plane \citep[e.g.][]{Obreschkow14,Fall18} they remained mostly unaffected by such processes. For a small number of galaxies, we identified processes which are able to considerably boost the spin of infalling galaxies during their infall via mergers with other objects and via transient encounters with stellar remnants within the cluster halo. These two processes are able to offset the numerical slow-down induced by numerical effects, such that at $z=0$ they end up within the top 5 per cent galaxies with the highest amount of ordered rotation within the simulation.
    
    Due the mass resolution of our simulations, which favours the modelling of a large number of galaxy clusters and their environments, our sample contains a limited kinematic distribution of galaxies when compared with observations \citep[e.g.][]{Brough17}. Thus, generalising these environmental constrains to galaxies with higher stellar angular momentum is not possible, as mass-loss processes during the infall of such galaxies in cluster environments might prove to play a more crucial role in their stellar angular momentum evolution. Future simulations, where the stellar mass resolution is increased and a smaller softening scale is used to better resolve the stellar component, for similar cosmological volumes, would help us extend our analysis to a wider range of galaxies with higher fractions of ordered rotation and obtain environmental constraints with far greater statistical significance.

\begin{acknowledgements}
    This work has been made possible by the `The Three Hundred' collaboration.\footnote{\url{https://www.the300-project.org}} The project has received financial support from the European Union's Horizon 2020 Research and Innovation programme under the Marie Sklodowskaw-Curie grant agreement number 734374, i.e. the LACEGAL project.

    The authors would like to thank the anonymous referee for their constructive feedback, which improved the quality of this paper. Additionally, they would like to thank The Red Espa\~{n}ola de Supercomputaci\'{o}n for granting us computing time at the MareNostrum Supercomputer of the BSC-CNS where most of the cluster simulations have been performed. Part of the computations with \code{GADGET-X} have also been performed at the `Leibniz-Rechenzentrum' with CPU time assigned to the Project `pr83li'.
    
    RM, AK, and GY would like to thank MINECO/FEDER (Spain) for financial support under research grants AYA2015-63819-P and MICIU/FEDER for financial support under research grant PGC2018-094975-C21. RM further acknowledges support from the International Centre for Radio Astronomy (ICRAR) for funding this project. AK further acknowledges support from the Spanish Red Consolider MultiDark FPA2017- 90566-REDC and thanks Joy Division for unknown pleasures. CL has received funding from the ARC Centre of Excellence for All Sky Astrophysics in 3 Dimensions (ASTRO 3D), through project number CE170100013. WC acknowledges the supported by the European Research Council under grant number 670193. SB acknowledges financial support from PRIN-MIUR 2015W7KAWC, the INFN INDARK grant, and the EU H2020 Research and Innovation Programme under the ExaNeSt project (Grant Agreement No. 671553). KD acknowledges support through ORIGINS, founded by the Deutsche Forschungsgemeinschaft (DFG, German Research Foundation) under Germany's Excellence Strategy - EXC-2094 - 390783311.
    
    The authors contributed to this paper in the following ways: RM, AK, FRP and CP formed the core team. RM analysed the data, produced the plots and wrote the paper along with AK, FRP, CL, WC, SB, and KD.  SB, KD, GM, and GY supplied the simulation data. All authors had the opportunity to provide comments on this work.
    
    This work was created by making use of the following software: \code{Python}, \code{Matplotlib} \citep{Matplotlib-Hunter07}, \code{Numpy} \citep{Numpy-vanDerWalt11}, \code{scipy} \citep{Scipy-Virtanen19}, \code{astropy} \citep{AstropyI,AstropyII}, and \code{pynbody} \citep{pynbody}.
\end{acknowledgements}



\bibliographystyle{aa}
\bibliography{archive}

\begin{thebibliography}{87}
\expandafter\ifx\csname natexlab\endcsname\relax\def\natexlab#1{#1}\fi

\bibitem[{{Abadi} {et~al.}(1999){Abadi}, {Moore}, \& {Bower}}]{Abadi99}
{Abadi}, M.~G., {Moore}, B., \& {Bower}, R.~G. 1999, \mnras, 308, 947

\bibitem[{{Arthur} {et~al.}(2017){Arthur}, {Pearce}, {Gray}, {Elahi}, {Knebe},
  {Beck}, {Cui}, {Cunnama}, {Dav{\'e}}, {February}, {Huang}, {Katz}, {Kay},
  {McCarthy}, {Murante}, {Perret}, {Power}, {Puchwein}, {Saro}, {Sembolini},
  {Teyssier}, \& {Yepes}}]{Arthur17}
{Arthur}, J., {Pearce}, F.~R., {Gray}, M.~E., {et~al.} 2017, \mnras, 464, 2027

\bibitem[{{Arthur} {et~al.}(2019){Arthur}, {Pearce}, {Gray}, {Knebe}, {Cui},
  {Elahi}, {Power}, {Yepes}, {Arth}, {De Petris}, {Dolag}, {Garratt-Smithson},
  {Old}, {Rasia}, \& {Stevens}}]{Arthur19}
{Arthur}, J., {Pearce}, F.~R., {Gray}, M.~E., {et~al.} 2019, \mnras, 484, 3968

\bibitem[{{Astropy Collaboration} {et~al.}(2018){Astropy Collaboration},
  {Price-Whelan}, {Sip{\H{o}}cz}, {G{\"u}nther}, {Lim}, {Crawford}, {Conseil},
  {Shupe}, {Craig}, {Dencheva}, {Ginsburg}, {Vand erPlas}, {Bradley},
  {P{\'e}rez-Su{\'a}rez}, {de Val-Borro}, {Aldcroft}, {Cruz}, {Robitaille},
  {Tollerud}, {Ardelean}, {Babej}, {Bach}, {Bachetti}, {Bakanov}, {Bamford},
  {Barentsen}, {Barmby}, {Baumbach}, {Berry}, {Biscani}, {Boquien}, {Bostroem},
  {Bouma}, {Brammer}, {Bray}, {Breytenbach}, {Buddelmeijer}, {Burke},
  {Calderone}, {Cano Rodr{\'\i}guez}, {Cara}, {Cardoso}, {Cheedella}, {Copin},
  {Corrales}, {Crichton}, {D'Avella}, {Deil}, {Depagne}, {Dietrich}, {Donath},
  {Droettboom}, {Earl}, {Erben}, {Fabbro}, {Ferreira}, {Finethy}, {Fox},
  {Garrison}, {Gibbons}, {Goldstein}, {Gommers}, {Greco}, {Greenfield},
  {Groener}, {Grollier}, {Hagen}, {Hirst}, {Homeier}, {Horton}, {Hosseinzadeh},
  {Hu}, {Hunkeler}, {Ivezi{\'c}}, {Jain}, {Jenness}, {Kanarek}, {Kendrew},
  {Kern}, {Kerzendorf}, {Khvalko}, {King}, {Kirkby}, {Kulkarni}, {Kumar},
  {Lee}, {Lenz}, {Littlefair}, {Ma}, {Macleod}, {Mastropietro}, {McCully},
  {Montagnac}, {Morris}, {Mueller}, {Mumford}, {Muna}, {Murphy}, {Nelson},
  {Nguyen}, {Ninan}, {N{\"o}the}, {Ogaz}, {Oh}, {Parejko}, {Parley}, {Pascual},
  {Patil}, {Patil}, {Plunkett}, {Prochaska}, {Rastogi}, {Reddy Janga},
  {Sabater}, {Sakurikar}, {Seifert}, {Sherbert}, {Sherwood-Taylor}, {Shih},
  {Sick}, {Silbiger}, {Singanamalla}, {Singer}, {Sladen}, {Sooley},
  {Sornarajah}, {Streicher}, {Teuben}, {Thomas}, {Tremblay}, {Turner},
  {Terr{\'o}n}, {van Kerkwijk}, {de la Vega}, {Watkins}, {Weaver}, {Whitmore},
  {Woillez}, {Zabalza}, \& {Astropy Contributors}}]{AstropyII}
{Astropy Collaboration}, {Price-Whelan}, A.~M., {Sip{\H{o}}cz}, B.~M., {et~al.}
  2018, \aj, 156, 123

\bibitem[{{Astropy Collaboration} {et~al.}(2013){Astropy Collaboration},
  {Robitaille}, {Tollerud}, {Greenfield}, {Droettboom}, {Bray}, {Aldcroft},
  {Davis}, {Ginsburg}, {Price-Whelan}, {Kerzendorf}, {Conley}, {Crighton},
  {Barbary}, {Muna}, {Ferguson}, {Grollier}, {Parikh}, {Nair}, {Unther},
  {Deil}, {Woillez}, {Conseil}, {Kramer}, {Turner}, {Singer}, {Fox}, {Weaver},
  {Zabalza}, {Edwards}, {Azalee Bostroem}, {Burke}, {Casey}, {Crawford},
  {Dencheva}, {Ely}, {Jenness}, {Labrie}, {Lim}, {Pierfederici}, {Pontzen},
  {Ptak}, {Refsdal}, {Servillat}, \& {Streicher}}]{AstropyI}
{Astropy Collaboration}, {Robitaille}, T.~P., {Tollerud}, E.~J., {et~al.} 2013,
  \aap, 558, A33

\bibitem[{{Bah{\'e}} {et~al.}(2017){Bah{\'e}}, {Barnes}, {Dalla Vecchia},
  {Kay}, {White}, {McCarthy}, {Schaye}, {Bower}, {Crain}, {Theuns}, {Jenkins},
  {McGee}, {Schaller}, {Thomas}, \& {Trayford}}]{Bahe17}
{Bah{\'e}}, Y.~M., {Barnes}, D.~J., {Dalla Vecchia}, C., {et~al.} 2017, \mnras,
  470, 4186

\bibitem[{{Bah{\'e}} \& {McCarthy}(2015)}]{Bahe15}
{Bah{\'e}}, Y.~M. \& {McCarthy}, I.~G. 2015, \mnras, 447, 969

\bibitem[{{Bah{\'e}} {et~al.}(2019){Bah{\'e}}, {Schaye}, {Barnes}, {Dalla
  Vecchia}, {Kay}, {Bower}, {Hoekstra}, {McGee}, \& {Theuns}}]{Bahe19}
{Bah{\'e}}, Y.~M., {Schaye}, J., {Barnes}, D.~J., {et~al.} 2019, \mnras, 485,
  2287

\bibitem[{{Bailin} \& {Steinmetz}(2005)}]{Bailin05}
{Bailin}, J. \& {Steinmetz}, M. 2005, \apj, 627, 647

\bibitem[{{Balogh} {et~al.}(2000){Balogh}, {Navarro}, \& {Morris}}]{Balogh00}
{Balogh}, M.~L., {Navarro}, J.~F., \& {Morris}, S.~L. 2000, \apj, 540, 113

\bibitem[{{Barnes} {et~al.}(2017){Barnes}, {Kay}, {Bah{\'e}}, {Dalla Vecchia},
  {McCarthy}, {Schaye}, {Bower}, {Jenkins}, {Thomas}, {Schaller}, {Crain},
  {Theuns}, \& {White}}]{Barnes17}
{Barnes}, D.~J., {Kay}, S.~T., {Bah{\'e}}, Y.~M., {et~al.} 2017, \mnras, 471,
  1088

\bibitem[{{Baxter} {et~al.}(2021){Baxter}, {Adhikari}, {Vega-Ferrero}, {Cui},
  {Chang}, {Jain}, \& {Knebe}}]{Baxter21}
{Baxter}, E.~J., {Adhikari}, S., {Vega-Ferrero}, J., {et~al.} 2021, arXiv
  e-prints, arXiv:2101.04179

\bibitem[{{Beck} {et~al.}(2016){Beck}, {Murante}, {Arth}, {Remus}, {Teklu},
  {Donnert}, {Planelles}, {Beck}, {F{\"o}rster}, {Imgrund}, {Dolag}, \&
  {Borgani}}]{Beck16}
{Beck}, A.~M., {Murante}, G., {Arth}, A., {et~al.} 2016, \mnras, 455, 2110

\bibitem[{{Behroozi} {et~al.}(2015){Behroozi}, {Knebe}, {Pearce}, {Elahi},
  {Han}, {Lux}, {Mao}, {Muldrew}, {Potter}, \& {Srisawat}}]{Behroozi15}
{Behroozi}, P., {Knebe}, A., {Pearce}, F.~R., {et~al.} 2015, \mnras, 454, 3020

\bibitem[{{Behroozi} {et~al.}(2014){Behroozi}, {Wechsler}, {Lu}, {Hahn},
  {Busha}, {Klypin}, \& {Primack}}]{Behroozi14}
{Behroozi}, P.~S., {Wechsler}, R.~H., {Lu}, Y., {et~al.} 2014, \apj, 787, 156

\bibitem[{{Brough} {et~al.}(2017){Brough}, {van de Sande}, {Owers},
  {d'Eugenio}, {Sharp}, {Cortese}, {Scott}, {Croom}, {Bassett}, {Bekki},
  {Bland-Hawthorn}, {Bryant}, {Davies}, {Drinkwater}, {Driver}, {Foster},
  {Goldstein}, {L{\'o}pez-S{\'a}nchez}, {Medling}, {Sweet}, {Taranu}, {Tonini},
  {Yi}, {Goodwin}, {Lawrence}, \& {Richards}}]{Brough17}
{Brough}, S., {van de Sande}, J., {Owers}, M.~S., {et~al.} 2017, \apj, 844, 59

\bibitem[{{Bullock} \& {Johnston}(2005)}]{Bullock05}
{Bullock}, J.~S. \& {Johnston}, K.~V. 2005, \apj, 635, 931

\bibitem[{{Capalbo} {et~al.}(2020){Capalbo}, {De Petris}, {De Luca}, {Cui},
  {Yepes}, {Knebe}, \& {Rasia}}]{Capalbo20}
{Capalbo}, V., {De Petris}, M., {De Luca}, F., {et~al.} 2020, \mnras
  [\eprint[arXiv]{2009.04565}]

\bibitem[{{Correa} {et~al.}(2017){Correa}, {Schaye}, {Clauwens}, {Bower},
  {Crain}, {Schaller}, {Theuns}, \& {Thob}}]{Correa17}
{Correa}, C.~A., {Schaye}, J., {Clauwens}, B., {et~al.} 2017, \mnras, 472, L45

\bibitem[{{Cortese} {et~al.}(2019){Cortese}, {van de Sande}, {Lagos},
  {Catinella}, {Davies}, {Croom}, {Brough}, {Bryant}, {Lawrence}, {Owers},
  {Richards}, {Sweet}, \& {Bland -Hawthorn}}]{Cortese19}
{Cortese}, L., {van de Sande}, J., {Lagos}, C.~P., {et~al.} 2019, \mnras, 485,
  2656

\bibitem[{{Cui} {et~al.}(2018){Cui}, {Knebe}, {Yepes}, {Pearce}, {Power},
  {Dave}, {Arth}, {Borgani}, {Dolag}, {Elahi}, {Mostoghiu}, {Murante}, {Rasia},
  {Stoppacher}, {Vega-Ferrero}, {Wang}, {Yang}, {Benson}, {Cora}, {Croton},
  {Sinha}, {Stevens}, {Vega-Mart{\'{\i}}nez}, {Arthur}, {Baldi}, {Ca{\~n}as},
  {Cialone}, {Cunnama}, {De Petris}, {Durando}, {Ettori}, {Gottl{\"o}ber},
  {Nuza}, {Old}, {Pilipenko}, {Sorce}, \& {Welker}}]{Cui18}
{Cui}, W., {Knebe}, A., {Yepes}, G., {et~al.} 2018, \mnras, 480, 2898

\bibitem[{{Cui} {et~al.}(2016){Cui}, {Power}, {Knebe}, {Kay}, {Sembolini},
  {Elahi}, {Yepes}, {Pearce}, {Cunnama}, {Beck}, {Dalla Vecchia}, {Dav{\'e}},
  {February}, {Huang}, {Hobbs}, {Katz}, {McCarthy}, {Murante}, {Perret},
  {Puchwein}, {Read}, {Saro}, {Teyssier}, \& {Thacker}}]{Cui16}
{Cui}, W., {Power}, C., {Knebe}, A., {et~al.} 2016, \mnras, 458, 4052

\bibitem[{{De Luca} {et~al.}(2020){De Luca}, {De Petris}, {Yepes}, {Cui},
  {Knebe}, \& {Rasia}}]{DeLuca20}
{De Luca}, F., {De Petris}, M., {Yepes}, G., {et~al.} 2020, arXiv e-prints,
  arXiv:2011.09002

\bibitem[{{Doroshkevich}(1970)}]{Doroshkevich70}
{Doroshkevich}, A.~G. 1970, Astrofizika, 6, 581

\bibitem[{{Dressler}(1980)}]{Dressler80}
{Dressler}, A. 1980, \apj, 236, 351

\bibitem[{{Elahi} {et~al.}(2016){Elahi}, {Knebe}, {Pearce}, {Power}, {Yepes},
  {Cui}, {Cunnama}, {Kay}, {Sembolini}, {Beck}, {Dav{\'e}}, {February},
  {Huang}, {Katz}, {McCarthy}, {Murante}, {Perret}, {Puchwein}, {Saro}, \&
  {Teyssier}}]{Elahi16}
{Elahi}, P.~J., {Knebe}, A., {Pearce}, F.~R., {et~al.} 2016, \mnras, 458, 1096

\bibitem[{{Errani} \& {Pe{\~n}arrubia}(2020)}]{Errani20}
{Errani}, R. \& {Pe{\~n}arrubia}, J. 2020, \mnras, 491, 4591

\bibitem[{{Fall}(1983)}]{Fall83}
{Fall}, S.~M. 1983, in IAU Symposium, Vol. 100, Internal Kinematics and
  Dynamics of Galaxies, ed. E.~{Athanassoula}, 391--398

\bibitem[{{Fall} \& {Romanowsky}(2018)}]{Fall18}
{Fall}, S.~M. \& {Romanowsky}, A.~J. 2018, \apj, 868, 133

\bibitem[{{Fujii} {et~al.}(2006){Fujii}, {Funato}, \& {Makino}}]{Fujii06}
{Fujii}, M., {Funato}, Y., \& {Makino}, J. 2006, \pasj, 58, 743

\bibitem[{{Fujita}(1998)}]{Fujita98}
{Fujita}, Y. 1998, \apj, 509, 587

\bibitem[{{Gill} {et~al.}(2004){Gill}, {Knebe}, \& {Gibson}}]{Gill04a}
{Gill}, S.~P.~D., {Knebe}, A., \& {Gibson}, B.~K. 2004, \mnras, 351, 399

\bibitem[{{Gunn} \& {Gott}(1972)}]{Gunn72}
{Gunn}, J.~E. \& {Gott}, J.~Richard, I. 1972, \apj, 176, 1

\bibitem[{{Haggar} {et~al.}(2020){Haggar}, {Gray}, {Pearce}, {Knebe}, {Cui},
  {Mostoghiu}, \& {Yepes}}]{Haggar20}
{Haggar}, R., {Gray}, M.~E., {Pearce}, F.~R., {et~al.} 2020, \mnras, 277

\bibitem[{{Hashimoto} \& {Oemler}(2000)}]{Hashimoto00}
{Hashimoto}, Y. \& {Oemler}, Augustus, J. 2000, \apj, 530, 652

\bibitem[{{Hayashi} {et~al.}(2003){Hayashi}, {Navarro}, {Taylor}, {Stadel}, \&
  {Quinn}}]{Hayashi03}
{Hayashi}, E., {Navarro}, J.~F., {Taylor}, J.~E., {Stadel}, J., \& {Quinn}, T.
  2003, \apj, 584, 541

\bibitem[{Hunter(2007)}]{Matplotlib-Hunter07}
Hunter, J.~D. 2007, Computing in Science \& Engineering, 9, 90

\bibitem[{{Jiang} \& {Binney}(2000)}]{Jiang00}
{Jiang}, I.-G. \& {Binney}, J. 2000, \mnras, 314, 468

\bibitem[{{Klimentowski} {et~al.}(2010){Klimentowski}, {{\L}okas}, {Knebe},
  {Gottl{\"o}ber}, {Martinez-Vaquero}, {Yepes}, \& {Hoffman}}]{Klimentowski10}
{Klimentowski}, J., {{\L}okas}, E.~L., {Knebe}, A.~e., {et~al.} 2010, \mnras,
  402, 1899

\bibitem[{{Klypin} {et~al.}(2016){Klypin}, {Yepes}, {Gottl{\"o}ber}, {Prada},
  \& {He{\ss}}}]{Klypin16}
{Klypin}, A., {Yepes}, G., {Gottl{\"o}ber}, S., {Prada}, F., \& {He{\ss}}, S.
  2016, \mnras, 457, 4340

\bibitem[{{Knebe} {et~al.}(2020){Knebe}, {G{\'a}mez-Mar{\'\i}n}, {Pearce},
  {Cui}, {Hoffmann}, {De Petris}, {Power}, {Haggar}, \& {Mostoghiu}}]{Knebe20}
{Knebe}, A., {G{\'a}mez-Mar{\'\i}n}, M., {Pearce}, F.~R., {et~al.} 2020,
  \mnras, 495, 3002

\bibitem[{{Knebe} {et~al.}(2006){Knebe}, {Power}, {Gill}, \&
  {Gibson}}]{Knebe06a}
{Knebe}, A., {Power}, C., {Gill}, S. P.~D., \& {Gibson}, B.~K. 2006, \mnras,
  368, 741

\bibitem[{{Knollmann} \& {Knebe}(2009)}]{Knollmann09}
{Knollmann}, S.~R. \& {Knebe}, A. 2009, \apjs, 182, 608

\bibitem[{{Kuchner} {et~al.}(2020){Kuchner}, {Arag{\'o}n-Salamanca}, {Pearce},
  {Gray}, {Rost}, {Mu}, {Welker}, {Cui}, {Haggar}, {Laigle}, {Knebe},
  {Kraljic}, {Sarron}, \& {Yepes}}]{Kuchner20}
{Kuchner}, U., {Arag{\'o}n-Salamanca}, A., {Pearce}, F.~R., {et~al.} 2020,
  \mnras, 494, 5473

\bibitem[{{Kuchner} {et~al.}(2021){Kuchner}, {Arag{\'o}n-Salamanca}, {Rost},
  {Pearce}, {Gray}, {Cui}, {Knebe}, {Rasia}, \& {Yepes}}]{Kuchner21}
{Kuchner}, U., {Arag{\'o}n-Salamanca}, A., {Rost}, A., {et~al.} 2021, \mnras,
  503, 2065

\bibitem[{{Lagos} {et~al.}(2018{\natexlab{a}}){Lagos}, {Schaye}, {Bah{\'e}},
  {Van de Sande}, {Kay}, {Barnes}, {Davis}, \& {Dalla Vecchia}}]{Lagos18b}
{Lagos}, C. d.~P., {Schaye}, J., {Bah{\'e}}, Y., {et~al.} 2018{\natexlab{a}},
  \mnras, 476, 4327

\bibitem[{{Lagos} {et~al.}(2018{\natexlab{b}}){Lagos}, {Stevens}, {Bower},
  {Davis}, {Contreras}, {Padilla}, {Obreschkow}, {Croton}, {Trayford},
  {Welker}, \& {Theuns}}]{Lagos18a}
{Lagos}, C. d.~P., {Stevens}, A. R.~H., {Bower}, R.~G., {et~al.}
  2018{\natexlab{b}}, \mnras, 473, 4956

\bibitem[{{Lagos} {et~al.}(2017){Lagos}, {Theuns}, {Stevens}, {Cortese},
  {Padilla}, {Davis}, {Contreras}, \& {Croton}}]{Lagos17}
{Lagos}, C. d.~P., {Theuns}, T., {Stevens}, A. R.~H., {et~al.} 2017, \mnras,
  464, 3850

\bibitem[{{Li} {et~al.}(2020){Li}, {Cui}, {Yang}, {Rasia}, {Dave}, {De Petris},
  {Knebe}, {Peacock}, {Pearce}, \& {Yepes}}]{Li20}
{Li}, Q., {Cui}, W., {Yang}, X., {et~al.} 2020, \mnras, 495, 2930

\bibitem[{{{\L}okas}(2020)}]{Lokas20}
{{\L}okas}, E.~L. 2020, \aap, 638, A133

\bibitem[{{Mazzarini} {et~al.}(2020){Mazzarini}, {Just}, {Macci{\`o}}, \&
  {Moetazedian}}]{Mazzarini20}
{Mazzarini}, M., {Just}, A., {Macci{\`o}}, A.~V., \& {Moetazedian}, R. 2020,
  \aap, 636, A106

\bibitem[{{Miller} {et~al.}(2020){Miller}, {van den Bosch}, {Green}, \&
  {Ogiya}}]{Miller20}
{Miller}, T.~B., {van den Bosch}, F.~C., {Green}, S.~B., \& {Ogiya}, G. 2020,
  arXiv e-prints, arXiv:2001.06489

\bibitem[{{Moore} {et~al.}(1996){Moore}, {Katz}, {Lake}, {Dressler}, \&
  {Oemler}}]{Moore96}
{Moore}, B., {Katz}, N., {Lake}, G., {Dressler}, A., \& {Oemler}, A. 1996,
  \nat, 379, 613

\bibitem[{{Moore} {et~al.}(1998){Moore}, {Lake}, \& {Katz}}]{Moore98}
{Moore}, B., {Lake}, G., \& {Katz}, N. 1998, \apj, 495, 139

\bibitem[{{Mostoghiu} {et~al.}(2021){Mostoghiu}, {Arthur}, {Pearce}, {Gray},
  {Knebe}, {Cui}, {Welker}, {Cora}, {Murante}, {Dolag}, \&
  {Yepes}}]{Mostoghiu21}
{Mostoghiu}, R., {Arthur}, J., {Pearce}, F.~R., {et~al.} 2021, \mnras, 501,
  5029

\bibitem[{{Mostoghiu} {et~al.}(2019){Mostoghiu}, {Knebe}, {Cui}, {Pearce},
  {Yepes}, {Power}, {Dave}, \& {Arth}}]{Mostoghiu19}
{Mostoghiu}, R., {Knebe}, A., {Cui}, W., {et~al.} 2019, \mnras, 483, 3390

\bibitem[{{Muldrew} {et~al.}(2011){Muldrew}, {Pearce}, \& {Power}}]{Muldrew11}
{Muldrew}, S.~I., {Pearce}, F.~R., \& {Power}, C. 2011, \mnras, 410, 2617

\bibitem[{{Obreschkow} \& {Glazebrook}(2014)}]{Obreschkow14}
{Obreschkow}, D. \& {Glazebrook}, K. 2014, \apj, 784, 26

\bibitem[{{Onions} {et~al.}(2012){Onions}, {Knebe}, {Pearce}, {Muldrew}, {Lux},
  {Knollmann}, {Ascasibar}, {Behroozi}, {Elahi}, {Han}, {Maciejewski},
  {Merch{\'a}n}, {Neyrinck}, {Ruiz}, {Sgr{\'o}}, {Springel}, \&
  {Tweed}}]{Onions12}
{Onions}, J., {Knebe}, A., {Pearce}, F.~R., {et~al.} 2012, \mnras, 2881

\bibitem[{{Park} {et~al.}(2007){Park}, {Choi}, {Vogeley}, {Gott}, {Blanton}, \&
  {SDSS Collaboration}}]{Park07}
{Park}, C., {Choi}, Y.-Y., {Vogeley}, M.~S., {et~al.} 2007, \apj, 658, 898

\bibitem[{{Pe{\~n}arrubia} {et~al.}(2008){Pe{\~n}arrubia}, {Navarro}, \&
  {McConnachie}}]{Penarrubia08b}
{Pe{\~n}arrubia}, J., {Navarro}, J.~F., \& {McConnachie}, A.~W. 2008, \apj,
  673, 226

\bibitem[{{Peebles}(1969)}]{Peebles69}
{Peebles}, P.~J.~E. 1969, \apj, 155, 393

\bibitem[{{Planck Collaboration} {et~al.}(2016){Planck Collaboration}, {Ade},
  {Aghanim}, {Arnaud}, {Ashdown}, {Aumont}, {Baccigalupi}, {Banday},
  {Barreiro}, {Bartlett}, {Bartolo}, {Battaner}, {Battye}, {Benabed},
  {Beno{\^\i}t}, {Benoit-L{\'e}vy}, {Bernard}, {Bersanelli}, {Bielewicz},
  {Bock}, {Bonaldi}, {Bonavera}, {Bond}, {Borrill}, {Bouchet}, {Boulanger},
  {Bucher}, {Burigana}, {Butler}, {Calabrese}, {Cardoso}, {Catalano},
  {Challinor}, {Chamballu}, {Chary}, {Chiang}, {Chluba}, {Christensen},
  {Church}, {Clements}, {Colombi}, {Colombo}, {Combet}, {Coulais}, {Crill},
  {Curto}, {Cuttaia}, {Danese}, {Davies}, {Davis}, {de Bernardis}, {de Rosa},
  {de Zotti}, {Delabrouille}, {D{\'e}sert}, {Di Valentino}, {Dickinson},
  {Diego}, {Dolag}, {Dole}, {Donzelli}, {Dor{\'e}}, {Douspis}, {Ducout},
  {Dunkley}, {Dupac}, {Efstathiou}, {Elsner}, {En{\ss}lin}, {Eriksen},
  {Farhang}, {Fergusson}, {Finelli}, {Forni}, {Frailis}, {Fraisse},
  {Franceschi}, {Frejsel}, {Galeotta}, {Galli}, {Ganga}, {Gauthier}, {Gerbino},
  {Ghosh}, {Giard}, {Giraud-H{\'e}raud}, {Giusarma}, {Gjerl{\o}w},
  {Gonz{\'a}lez-Nuevo}, {G{\'o}rski}, {Gratton}, {Gregorio}, {Gruppuso},
  {Gudmundsson}, {Hamann}, {Hansen}, {Hanson}, {Harrison}, {Helou},
  {Henrot-Versill{\'e}}, {Hern{\'a}ndez-Monteagudo}, {Herranz}, {Hildebrand t},
  {Hivon}, {Hobson}, {Holmes}, {Hornstrup}, {Hovest}, {Huang}, {Huffenberger},
  {Hurier}, {Jaffe}, {Jaffe}, {Jones}, {Juvela}, {Keih{\"a}nen}, {Keskitalo},
  {Kisner}, {Kneissl}, {Knoche}, {Knox}, {Kunz}, {Kurki-Suonio}, {Lagache},
  {L{\"a}hteenm{\"a}ki}, {Lamarre}, {Lasenby}, {Lattanzi}, {Lawrence}, {Leahy},
  {Leonardi}, {Lesgourgues}, {Levrier}, {Lewis}, {Liguori}, {Lilje},
  {Linden-V{\o}rnle}, {L{\'o}pez-Caniego}, {Lubin}, {Mac{\'\i}as-P{\'e}rez},
  {Maggio}, {Maino}, {Mandolesi}, {Mangilli}, {Marchini}, {Maris}, {Martin},
  {Martinelli}, {Mart{\'\i}nez-Gonz{\'a}lez}, {Masi}, {Matarrese}, {McGehee},
  {Meinhold}, {Melchiorri}, {Melin}, {Mendes}, {Mennella}, {Migliaccio},
  {Millea}, {Mitra}, {Miville-Desch{\^e}nes}, {Moneti}, {Montier}, {Morgante},
  {Mortlock}, {Moss}, {Munshi}, {Murphy}, {Naselsky}, {Nati}, {Natoli},
  {Netterfield}, {N{\o}rgaard-Nielsen}, {Noviello}, {Novikov}, {Novikov},
  {Oxborrow}, {Paci}, {Pagano}, {Pajot}, {Paladini}, {Paoletti}, {Partridge},
  {Pasian}, {Patanchon}, {Pearson}, {Perdereau}, {Perotto}, {Perrotta},
  {Pettorino}, {Piacentini}, {Piat}, {Pierpaoli}, {Pietrobon}, {Plaszczynski},
  {Pointecouteau}, {Polenta}, {Popa}, {Pratt}, {Pr{\'e}zeau}, {Prunet},
  {Puget}, {Rachen}, {Reach}, {Rebolo}, {Reinecke}, {Remazeilles}, {Renault},
  {Renzi}, {Ristorcelli}, {Rocha}, {Rosset}, {Rossetti}, {Roudier},
  {Rouill{\'e} d'Orfeuil}, {Rowan-Robinson}, {Rubi{\~n}o-Mart{\'\i}n},
  {Rusholme}, {Said}, {Salvatelli}, {Salvati}, {Sandri}, {Santos},
  {Savelainen}, {Savini}, {Scott}, {Seiffert}, {Serra}, {Shellard}, {Spencer},
  {Spinelli}, {Stolyarov}, {Stompor}, {Sudiwala}, {Sunyaev}, {Sutton},
  {Suur-Uski}, {Sygnet}, {Tauber}, {Terenzi}, {Toffolatti}, {Tomasi},
  {Tristram}, {Trombetti}, {Tucci}, {Tuovinen}, {T{\"u}rler}, {Umana},
  {Valenziano}, {Valiviita}, {Van Tent}, {Vielva}, {Villa}, {Wade}, {Wandelt},
  {Wehus}, {White}, {White}, {Wilkinson}, {Yvon}, {Zacchei}, \&
  {Zonca}}]{Planck15}
{Planck Collaboration}, {Ade}, P.~A.~R., {Aghanim}, N., {et~al.} 2016, \aap,
  594, A13

\bibitem[{{Planelles} {et~al.}(2017){Planelles}, {Fabjan}, {Borgani},
  {Murante}, {Rasia}, {Biffi}, {Truong}, {Ragone-Figueroa}, {Granato}, {Dolag},
  {Pierpaoli}, {Beck}, {Steinborn}, \& {Gaspari}}]{Planelles17}
{Planelles}, S., {Fabjan}, D., {Borgani}, S., {et~al.} 2017, \mnras, 467, 3827

\bibitem[{{Pontzen} {et~al.}(2013){Pontzen}, {Ro{\v s}kar}, {Stinson}, \&
  {Woods}}]{pynbody}
{Pontzen}, A., {Ro{\v s}kar}, R., {Stinson}, G., \& {Woods}, R. 2013, {pynbody:
  N-Body/SPH analysis for python}, Astrophysics Source Code Library

\bibitem[{{Power} {et~al.}(2018){Power}, {Elahi}, {Welker}, {Knebe}, {Pearce},
  {Yepes}, {Dave}, {Kay}, {McCarthy}, {Puchwein}, {Borgani}, {Cunnama}, {Cui},
  \& {Schaye}}]{Power18}
{Power}, C., {Elahi}, P.~J., {Welker}, C., {et~al.} 2018, arXiv e-prints,
  arXiv:1810.00534

\bibitem[{{Rasia} {et~al.}(2015){Rasia}, {Borgani}, {Murante}, {Planelles},
  {Beck}, {Biffi}, {Ragone-Figueroa}, {Granato}, {Steinborn}, \&
  {Dolag}}]{Rasia15}
{Rasia}, E., {Borgani}, S., {Murante}, G., {et~al.} 2015, \apjl, 813, L17

\bibitem[{{Recchi}(2014)}]{Recchi14}
{Recchi}, S. 2014, Advances in Astronomy, 2014, 750754

\bibitem[{{Romanowsky} \& {Fall}(2012)}]{Romanowsky12}
{Romanowsky}, A.~J. \& {Fall}, S.~M. 2012, \apjs, 203, 17

\bibitem[{{Rost} {et~al.}(2021){Rost}, {Kuchner}, {Welker}, {Pearce},
  {Stasyszyn}, {Gray}, {Cui}, {Dave}, {Knebe}, {Yepes}, \& {Rasia}}]{Rost21}
{Rost}, A., {Kuchner}, U., {Welker}, C., {et~al.} 2021, \mnras, 502, 714

\bibitem[{{Sales} {et~al.}(2010){Sales}, {Navarro}, {Schaye}, {Dalla Vecchia},
  {Springel}, \& {Booth}}]{Sales10}
{Sales}, L.~V., {Navarro}, J.~F., {Schaye}, J., {et~al.} 2010, \mnras, 409,
  1541

\bibitem[{{Schulze} {et~al.}(2018){Schulze}, {Remus}, {Dolag}, {Burkert},
  {Emsellem}, \& {van de Ven}}]{Schulze18}
{Schulze}, F., {Remus}, R.-S., {Dolag}, K., {et~al.} 2018, \mnras, 480, 4636

\bibitem[{{Sembolini} {et~al.}(2016{\natexlab{a}}){Sembolini}, {Elahi},
  {Pearce}, {Power}, {Knebe}, {Kay}, {Cui}, {Yepes}, {Beck}, {Borgani},
  {Cunnama}, {Dav{\'e}}, {February}, {Huang}, {Katz}, {McCarthy}, {Murante},
  {Newton}, {Perret}, {Puchwein}, {Saro}, {Schaye}, \&
  {Teyssier}}]{Sembolini16b}
{Sembolini}, F., {Elahi}, P.~J., {Pearce}, F.~R., {et~al.} 2016{\natexlab{a}},
  \mnras, 459, 2973

\bibitem[{{Sembolini} {et~al.}(2016{\natexlab{b}}){Sembolini}, {Yepes},
  {Pearce}, {Knebe}, {Kay}, {Power}, {Cui}, {Beck}, {Borgani}, {Dalla Vecchia},
  {Dav{\'e}}, {Elahi}, {February}, {Huang}, {Hobbs}, {Katz}, {Lau}, {McCarthy},
  {Murante}, {Nagai}, {Nelson}, {Newton}, {Perret}, {Puchwein}, {Read}, {Saro},
  {Schaye}, {Teyssier}, \& {Thacker}}]{Sembolini16a}
{Sembolini}, F., {Yepes}, G., {Pearce}, F.~R., {et~al.} 2016{\natexlab{b}},
  \mnras, 457, 4063

\bibitem[{{Smith} {et~al.}(2016){Smith}, {Choi}, {Lee}, {Rhee},
  {Sanchez-Janssen}, \& {Yi}}]{Smith16}
{Smith}, R., {Choi}, H., {Lee}, J., {et~al.} 2016, \apj, 833, 109

\bibitem[{{Smith} {et~al.}(2010){Smith}, {Davies}, \& {Nelson}}]{Smith10}
{Smith}, R., {Davies}, J.~I., \& {Nelson}, A.~H. 2010, \mnras, 405, 1723

\bibitem[{{Smith} {et~al.}(2015){Smith}, {S{\'a}nchez-Janssen}, {Beasley},
  {Cand lish}, {Gibson}, {Puzia}, {Janz}, {Knebe}, {Aguerri}, {Lisker},
  {Hensler}, {Fellhauer}, {Ferrarese}, \& {Yi}}]{Smith15}
{Smith}, R., {S{\'a}nchez-Janssen}, R., {Beasley}, M.~A., {et~al.} 2015,
  \mnras, 454, 2502

\bibitem[{{Teklu} {et~al.}(2015){Teklu}, {Remus}, {Dolag}, {Beck}, {Burkert},
  {Schmidt}, {Schulze}, \& {Steinborn}}]{Teklu15}
{Teklu}, A.~F., {Remus}, R.-S., {Dolag}, K., {et~al.} 2015, \apj, 812, 29

\bibitem[{{Valtonen} {et~al.}(1990){Valtonen}, {Valtaoja}, {Sundelius},
  {Donner}, \& {Byrd}}]{Valtonen90}
{Valtonen}, M.~J., {Valtaoja}, L., {Sundelius}, B., {Donner}, K.~J., \& {Byrd},
  G.~G. 1990, Celestial Mechanics and Dynamical Astronomy, 48, 95

\bibitem[{{van den Bosch}(2017)}]{vanDenBosch17}
{van den Bosch}, F.~C. 2017, \mnras, 468, 885

\bibitem[{{van der Walt} {et~al.}(2011){van der Walt}, {Colbert}, \&
  {Varoquaux}}]{Numpy-vanDerWalt11}
{van der Walt}, S., {Colbert}, S.~C., \& {Varoquaux}, G. 2011, Computing in
  Science Engineering, 13, 22

\bibitem[{{Vega-Ferrero} {et~al.}(2021){Vega-Ferrero}, {Dana}, {Diego},
  {Yepes}, {Cui}, \& {Meneghetti}}]{VegaFerrero21}
{Vega-Ferrero}, J., {Dana}, J.~M., {Diego}, J.~M., {et~al.} 2021, \mnras, 500,
  247

\bibitem[{{Virtanen} {et~al.}(2019){Virtanen}, {Gommers}, {Oliphant},
  {Haberland}, {Reddy}, {Cournapeau}, {Burovski}, {Peterson}, {Weckesser},
  {Bright}, {van der Walt}, {Brett}, {Wilson}, {Jarrod Millman}, {Mayorov},
  {Nelson}, {Jones}, {Kern}, {Larson}, {Carey}, {Polat}, {Feng}, {Moore}, {Vand
  erPlas}, {Laxalde}, {Perktold}, {Cimrman}, {Henriksen}, {Quintero}, {Harris},
  {Archibald}, {Ribeiro}, {Pedregosa}, {van Mulbregt}, \&
  {Contributors}}]{Scipy-Virtanen19}
{Virtanen}, P., {Gommers}, R., {Oliphant}, T.~E., {et~al.} 2019, arXiv
  e-prints, arXiv:1907.10121

\bibitem[{{Wang} {et~al.}(2018){Wang}, {Pearce}, {Knebe}, {Yepes}, {Cui},
  {Power}, {Arth}, {Gottl{\"o}ber}, {De Petris}, {Brown}, \& {Feng}}]{Wang18}
{Wang}, Y., {Pearce}, F., {Knebe}, A., {et~al.} 2018, \apj, 868, 130

\bibitem[{{Wang} {et~al.}(2016){Wang}, {Pearce}, {Knebe}, {Schneider},
  {Srisawat}, {Tweed}, {Jung}, {Han}, \& {et al.}}]{Wang16}
{Wang}, Y., {Pearce}, F.~R., {Knebe}, A., {et~al.} 2016, \mnras, 459, 1554

\bibitem[{{White}(1984)}]{White84}
{White}, S.~D.~M. 1984, \apj, 286, 38

\bibitem[{{White} \& {Rees}(1978)}]{White78}
{White}, S.~D.~M. \& {Rees}, M.~J. 1978, \mnras, 183, 341

\end{thebibliography}



\appendix


\end{document}